\documentclass[a4paper]{article}
\usepackage{graphicx}
 \newcommand{\be}{\begin{equation}}
\newcommand{\ee}{\end{equation}}  
\def\fun#1#2{\lower3.6pt\vbox{\baselineskip0pt\lineskip.9pt
\ialign{$\mathsurround=0pt#1\hfil ##\hfil$\crcr#2\crcr\sim\crcr}}}

\title {
Novel solutions of RG equations for $\alpha(s)$ and
$\beta(\alpha)$ in the large $N_c$ limit}
\author{
Yu.A. Simonov}
\date{}
\begin{document}

\maketitle
\begin{abstract}

General solution of RG equations in the framework of background
perturbation theory is written down in the large $N_c$ limit. A
simplified (model) approximation to the general solution is
suggested which allows to write $\beta(\alpha)$ and
$\alpha(\beta)$ to any loop order. The resulting $\alpha_B(Q^2)$
coincides asymptotically at large $|Q^2|$ with standard (free)
$\alpha_s$, saturates at small $Q^2 \geq 0$ and has poles at
time-like $Q^2$ in agreement with analytic properties of physical
amplitudes in the large $N_c$ limit.

\end{abstract}

\newpage


1. The running $\alpha_s$ of Standard Perturbation Theory (SPT) has
unphysical singularities (Landau ghost pole and cuts) in the
Euclidean region; which contradict analiticity of physical
amplitudes and are not observed  in lattice calculations of
$\alpha_s(Q^2)$ \cite{1}.

As it was shown in \cite{2,3} the Background Perturbation Theory
(BPT) is free of these defects and the coupling constant
$\alpha_B$ of BPT displays an important property of IR freezing
(saturation) with $\alpha_B(Q^2=0) \approx 0.5$ \cite{2,3}. This
behaviour of $\alpha_B$ is well confirmed by experimental data on
spin splitting of quarkonia levels \cite{4,5} and by lattice data
\cite{6}.

Analytic properties of $\alpha_B(Q^2)$ in the $Q^2$ plane were not
studied heretofore, and is the topic of the present letter. To
this end the large $N_c$ limit is exploited which ensures that any
physical amplitude has only simple poles \cite{7}. This requires
that $\alpha_B(Q^2)$ should be a meromorphic function of $Q^2$.

To find it explicitly we study below the general solution of RG
equations and suggest a simple model approximation to it, yielding
$\beta(\alpha)$ to all orders and explicit form of
$\alpha_B(Q^2)$.

Resulting $\alpha_B(Q^2)$ can be compared with $\alpha_s(Q^2)$ of
SPT and coincides with the latter asymptotically at large $Q^2$.


2. Separating background field $B_\mu$ in the total gluonic field
$A_\mu = B_\mu +a_\mu$ and expanding in $ga_\mu$ one gets the BPT
series, where coefficients depend both on external momenta and the
background field $g B_\mu$, which turns out to be RG invariant, if
background gauge is chosen \cite{8}. Correspondingly $\alpha_B
\equiv {g^2}/{4\pi}$ satisfies the standard RG equations,
where momenta and averaged background characteristic ${\rm
tr}<F_{\mu_1 \nu_1}(B) F_{\mu_2 \nu_2}(B) ... F_{\mu_n \nu_n}(B)>$
enter together and will be denoted $\{{\cal P}_i^2\}$.

One starts with the standard definition
\begin{equation}
\frac{d\alpha_B(\mu)}{d\ln \mu} = \beta(\alpha_B) \label{1}
\end{equation}
and represents $\beta(\alpha_B)$ in the form
\begin{equation}
\beta(\alpha) = -\frac{\beta_0}{2\pi}
\frac{\alpha^2}{[1-\frac{\beta_1}{2\pi\beta_0}\varphi^{\prime}(\frac{1}{\alpha})]}
\label{2}
\end{equation}
where $\varphi(z)$ is an arbitrary function with some conditions
to be imposed below, and prime means derivative.

Solving (\ref{1}) with the help of (\ref{2}) one has
\begin{equation}
\alpha_B = \frac{4\pi}{\beta[\ln\mu^2 +\chi +
\frac{2\beta_1}{\beta_0^2}\varphi (\frac{1}{\alpha_B})]} \label{3}
\end{equation}

Here $\chi$ is an arbitrary dimensionless function of $\{{\cal
P}_i^2\}$, while $\varphi$ is partly fixed by few coefficients of
expansion of $\beta(\alpha)$,
\begin{equation}
\beta(\alpha) = -\frac{\beta_0}{2\pi}\alpha^2
-\frac{\beta_1}{4\pi^2}\alpha^3 -\frac{\beta_2}{64\pi^3}\alpha^4 -
\frac{\beta_3\alpha^5}{(4\pi)^4} \label{4}
\end{equation}
(Note that $\beta_i$ are defined as in \cite{9}, for $n_f=0,
N_c=3$ and $\overline{MS}$ scheme
$\beta_0=11,~~\beta_1=51,~~\beta_2=2857,~~\beta_3=58486$.)

At this point we shall use the large $N_c$ limit and require that
$\alpha_B(Q^2)$ be a meromorphic function of $Q^2$. Consider e.g.
the process $e^+ e^-$ into hadrons, and the photon self-energy
part $\Pi(Q^2)$, which has a pole expansion \cite{3}
\begin{equation}
\Pi(Q^2, \alpha_B) = \sum^\infty_{n=0}\frac{C_n(\alpha_B)}{Q^2
+M_n^2(\alpha_B)} + {\rm subtractions} \label{5}
\end{equation}

Expanding in $\alpha_B$ and doing a partial summation (equivalent
to the summation of leading logarithms) one obtains a running
$\alpha_B(Q^2)$ which has poles at some generally speaking,
shifted positions as compared to the poles of the lowest order --
$M_n^2(0)$. The latter are proportional to $n$; $M_n^2(0)=m_0^2 n
+M_0^2$; $n=0,1,2,...$, $m_0^2 =4\pi\sigma$, $C_n(0)=m_0^2$
\cite{3}. The lowest (partonic) approximation thus reduces to
$\Pi(Q^2, 0) \sim \psi(\frac{Q^2 +M_0^2}{m^2})$, where
$\psi(z)=\frac{\Gamma^{\prime}(z)}{\Gamma(z)}$.

To specify the form of the function $\varphi(\frac{1}{\alpha_B})$
and $\chi(Q^2)$, we shall use two requirements:
\begin{description}
\item{i)} the correspondence principle with the SPT, which tells
that for $Q^2 \gg \kappa^2$, where $\kappa^2$ is the scale of
nonperturbative vacuum fields, our solution should coincide with
that of SPT.
\item{ii)} $\alpha_B(Q^2)$ should be meromorphic function of
$Q^2$, and hence both $\chi$ and $\varphi$ should be meromorphic
functions of their arguments ($Q^2$ and $\frac{1}{\alpha_B}$
respectively).
\end{description}

The first requirement means that coefficients $C_n$ in the sum
over poles should have finite limit $C_\infty$, i.e. one can
represent this sum as the Euler function $\psi(z) =
\frac{\Gamma^\prime(z)}{\Gamma(z)}$ plus a sum with fast
decreasing coefficients, which we write symbolically as a finite
sum. Thus the general solution satisfying i) and ii) can be
written as
\begin{equation}
\ln\mu^2 +\chi = \ln\frac{m^2}{\Lambda^2} +\psi(\frac{Q^2
+M_0^2}{m^2}) +\sum_{n=1}^{N_1} \frac{b_n}{Q^2 +m_n^2} \label{6}
\end{equation}
\begin{equation}
\varphi(\frac{1}{\alpha_B})= \psi(\frac{\lambda}{\alpha_B}+\Delta)
+P\left(\frac{1}{\alpha_B}\right)  \label{7}
\end{equation}
where $\Delta$ is some constant and $\lambda =
{4\pi}/{\beta_0}$. Note that our solution (\ref{6}), (\ref{7})
satisfies (\ref{1}) and reproduces the first two coefficients in
the expansion (\ref{4}) when all $b_n = 0$ and $P\equiv 0$, while
our $\beta_2,\beta_3$ appear to be smaller; however the latter are
scheme-dependent. It is easy to reproduce $\beta_2$, $\beta_3$ by
choosing in (\ref{7}) $P\neq 0$.  This choice of
$\varphi(\frac{1}{\alpha_B})$ leads to a small numerical (around
2\%) change in $\alpha_B$ as compared to the minimal choice
(\ref{8}) for all $Q^2 \leq 0$.

3. In what follows we confine ourselves to the minimal model of
(\ref{6}), (\ref{7}) with all $b_n$, and $P$ identically zero,
which yields
\begin{equation}
\alpha_B =\frac{4\pi}{\beta_0} \left[\ln\frac{m^2}{\Lambda^2}
+\psi (\frac{Q^2 +M_0^2}{m^2}) +\frac{2\beta_1}{\beta_0^2}\psi
(\frac{\lambda}{\alpha_B} +\Delta)\right]^{-1} \label{8}
\end{equation}
 Eq.(\ref{8}) defines $\alpha_B$ as a meromorphic function of
$Q^2$ in the whole $Q^2$ complex plane.

Moreover, Eq.(\ref{8}) defines $\alpha_B$ to all orders in the
``loop expansion'' obtained by iteration of the last term in the
denominator of (\ref{8}) and will be compared below to the SPT
loop expansion. Consider e.g. large $Q^2$, $\frac{Q^2 +M_0^2}{m^2}
\gg 1$. Then using the asymptotic expansion for $z \to \infty$
\begin{equation} \psi(z) = \ln z -\frac{1}{2z} -\sum_{k=1}^\infty
\frac{B_{2k}}{2kz^{2k}} \;\; , \label{9}
\end{equation}
where $B_n$ are Bernoulli numbers, one obtains to the lowest order
\begin{equation}
\alpha_B^{(0)} =\frac{4\pi}{\beta_0 \ln(\frac{Q^2
+M_0^2}{\Lambda^2})} \label{10}
\end{equation}
which explicitly shows the AF behaviour for large $Q^2$ and
absence of Landau pole (for $M_0^2 >\Lambda^2$).

The forms (\ref{2}) and (\ref{8}) admit in general poles of
$\alpha_B(Q^2)$ for $Q^2 \geq 0$ and poles of $\beta(\alpha_B)$
for $\alpha_B \geq 0$, which would be unphysical. The latter poles
are defined by equation $1=\frac{\beta_1
\lambda}{2\pi\beta_0}\psi^{\prime}(\frac{\lambda}{\alpha}+\Delta)$,
and are excluded from the region $\alpha \geq 0$ when $\Delta >
\Delta_0 = 1.255$. This latter condition also excludes
singularities of $\alpha_B(Q^2)$ (\ref{8}), at $Q^2>0$.  At the
same time $\beta(\alpha)$ has a point of condensing zeros on the
negative side of $\alpha =0$, and $\beta(\alpha)$ is monotonically
decreasing when $\alpha \to \infty$, as it is shown in Table 1.
Note that a pole of $\beta(\alpha)$ at $\alpha=\alpha^*>0$ leads
to a situation when real solutions $\alpha(q)$ exist only for
$\alpha\leq\alpha^*,Q>Q^*$.

\begin{figure}[!ht]
\begin{picture}(320,220)
\put(0,15){\includegraphics{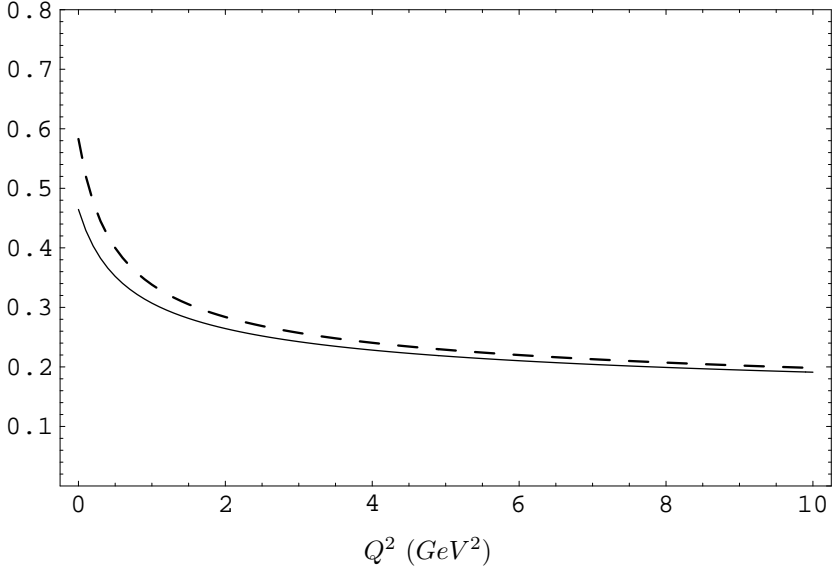}}
\put(0,90){\rotatebox{90}{}}
\put(140,1){$Q^2 ~(GeV^2)$}
\put(100,120){}
\end{picture}
\caption{Exact solution of Eq.(8) with $\Delta = 1.5$ as a function of $s \equiv Q^2$ (solid
line), and exact solution for $\alpha_B$ from the extended model, Eq.(3) with $\varphi$
from (21) and $\eta =3, \Delta = 1.5$ which reproduces all known $\beta_n$ coefficients
(broken line).}
\label{Fig.1}
\end{figure}

4. Although $\alpha_B(Q^2)$ does not have poles for $\Delta>\Delta_0$
in the Euclidean region, $Q^2>0$, it has an infinite number of
poles for $Q^2<0$.

These poles are given by zeros of the denominator in (\ref{8}),
which can be written as
\be
-Q^2\equiv s_n= s_n^{(0)}- \delta_n, ~~ s_n^{(0)}\equiv M^2_n=
M_0^2+nm^2\label{11} \ee and $\delta$ is then found from (\ref{8})
to be \cite{10a}
\be
\delta_n^{-1}\cong \ln \frac{M^2_n}{\Lambda^2}.\label{12} \ee

Hence poles of $\alpha_B$ are shifted as compared with poles of
$\psi\left(\frac{Q^2+M^2_0}{m^2}\right)$, the latter describe the
physical states of (double) hybrids - the background analog of
gluon loops renormalizing $\alpha_s$ in SPT. In the vicinity of
the pole $\alpha_B$ has the form \cite{10a}
\be
\alpha_B(s\approx s_n-\delta_n)
=\frac{(4\pi/\beta_0-\psi'(\Delta))}{\psi'(z_n)(z-z_n)},~~z=\frac{M_0^2-s}{m^2}
\label{13} \ee

Hence perturbative  series of BPT is a sum  over poles of
$\alpha_B$ and nonperturbative poles  of $\Pi(Q^2,0)$ in (\ref{5})
(dual to partonic contribution), all poles being in the region
$s>M^2_0-\delta_0>0$ and this situation is in accord with the
large $N_c$ analytic properties. Now one has to show that the
"physically averaged value" of $\alpha_B(s)$ corresponds to the AF
behaviour, as it is seen e.g. in the $e^+e^-$ annihilation. To
this end one   one can choose two equivalent procedures: i) to
introduce the  width of the pole in $\alpha_B(s)$, or ii)
equivalently to consider the BPT series at the shifted value of
$s, s=\bar s(1+i\gamma)$ (see e.g.\cite{10}).

In what follows we shall exploit the second path. Considering
${\rm Im}\alpha_B$ above the real axis, at $\bar s(1+i\gamma)$ with
$\bar s$ large and $\gamma$ fixed $\gamma\ll \pi$, allows to use
asymptotics (\ref{10}) (in one loop order) with the result
\be
{\rm Im} \alpha (\bar s
(1+i\gamma))\approx\frac{4\pi(\pi-\gamma)}{b_0[\ln^2|\frac{M^2_0-\bar
s}{\Lambda^2}|+(\pi-\gamma)^2]} \label{14}
\ee

Another form of discontinuity results from the expansion of the
Adler function $D(Q^2)=Q^2\frac{d\Pi(Q^2)}{d\ln Q^2},$ from which
the hadronic ratio $R(s)$ can be obtained as \cite{11}.
\be
R(s)=\frac{1}{2\pi i}\int^{-s+i\varepsilon}_{-s-i\varepsilon}
D(\sigma) \frac{d\sigma}{\sigma} =\frac{1}{2\pi i} \int_{C(s)}
D(\sigma) \frac{d\sigma}{\sigma} \label{15} \ee

Here $C(s)$ is the circle of radius  $s$ around the origin,
comprising all singularities of $D(\sigma)$ in Euclidean region.

Similarly to \cite{11}
 one can define the operation $\Phi\{\frac{\alpha_B}{\pi}\}$
 transforming the known perturbative series for $R(s)$;
 \be
 R(s)=N_c\sum_q e^2_q\left[1+\bar \Phi\left\{\frac{\alpha_B}{\pi}\right\}+d_2\bar
 \Phi\left\{\left(\frac{\alpha_B}{\pi}\right)^2\right\}\right]\label{16}
 \ee
 where
 \be
 \bar\Phi \left\{\left(\frac{\alpha_B}{\pi}\right)^k\right\} =\frac{1}{2\pi i}
\int_{C(s)}\left(\frac{ \alpha_B(\sigma)}{\pi}\right)^k
\frac{d\sigma}{\sigma} \label{17}\ee

We shall consider the situation, when one can use the logarithmic
asymptotics (\ref{10}) for $\alpha_B$ in the Minkowskian region.
This happens in the physical situation described above:  when
poles have imaginary parts (widths of resonances), or equivalently
when one consider effective $\bar R(s)=\frac{1}{2i}
(\Pi(s(1+i\gamma))-\Pi(s(1-i\gamma))$ with fixed $\gamma$ (so that
requirement $|{\rm arg}(z=\frac{M_0^2-s}{m^2})|<\pi$ is fulfilled).

In this case one obtains, inserting in the integral (\ref{17}) the
asymptotic form (\ref{10}) (see \cite{10a} for details)
\be
\Phi\left\{\frac{\bar\alpha_B }{\pi}\right\}
\cong\frac{4}{b_0}\left \{ \frac{1}{\pi} {\rm arctg} \frac{\pi}{\ln
s/\Lambda^2} + \frac{2M^2_0}{s \ln^3 s/\Lambda^2} +
O\left(\frac{M^2_0}{s \ln^4 s/\Lambda^2}\right)\right\} \label{18}
\ee

5. One can now compare the power (``loop expansion'') of $\alpha_B$
in (\ref{8}) in powers of $\alpha_B^{(0)}$ (\ref{10}) with the SPT
expression for $\alpha_s$ \cite{9, 12}. To this end we shall write
it as follows:
\begin{equation}
\alpha_B =\alpha_B^{(0)} \left\{1-\frac{2\beta_1 \ln L}{\beta_0^2
L} +\frac{4\beta_1^2}{\beta_0^4 L^2} \left[(\ln L - \frac{1}{2})^2
+b\right] \right\} \label{19}
\end{equation}
where $L\equiv \ln(\frac{Q^2 +M_0^2}{\Lambda^2})$ and for $n_f =0,
\Delta = 1.5$
\begin{equation}
b =-\frac{\beta_0^2}{2\beta_1}(\Delta -\frac{1}{2}) -\frac{1}{4} =
-1.436  \label{20}
\end{equation}

This should be compared with the $\overline{MS}$ value of SPT,
$b_{\overline{MS}} = 0.26$ ($n_f =0$).

To bring our theoretical expressions for $\beta_2$, $\beta_3$ in
agreement with the computed $\overline{MS}$ values (see \cite{9}
for the corresponding references), one should generalize the
minimal model considered above by adding a term to
$\varphi(\frac{1}{\alpha})$:
\begin{equation}
\varphi^{\prime}(\frac{1}{\alpha}) =
\lambda\psi^\prime(\frac{\lambda}{\alpha}+\Delta)
+\frac{\tilde{C}\alpha^2}{1+\alpha^2\eta^2} \label{21}
\end{equation}
Choosing $\tilde{C} = 1.253$ and arbitrary real $\eta$, one
obtains both $\beta_2$, $\beta_3$ in close agreement (within 1\%)
with $\overline{MS}$ values. In this case also $b$ in (\ref{20})
gets contribution from the new term and becomes $\tilde{b} =
b+\frac{2\pi\beta_0}{\beta_1}\tilde{C} = 0.263$, which agrees with
the SPT \cite{9}. We study numerically both minimal model and the
corrected one as explained above and present the results  for
$\beta(\alpha)$ in Table 1 and for $\alpha_B(Q^2)$ in Table 2.

From Table 1 it is clear that nonperturbative $\beta(\alpha)$,
namely $\beta_{min}(\alpha)$, calculated as in (\ref{2}),
(\ref{7}) with $P\equiv 0$ and $\beta_{ext}(\alpha)$ calculated as
in (\ref{2}), (\ref{21}), both decrease more moderately (as
(-const $\alpha^2$)) at $\alpha>1$ as compared to purely
perturbative 3 loop expression (\ref{4}). This might  indicate
that the series (\ref{4}) is asymptotic and should be cut off
after few first terms (note that $\beta_3$ is larger than previous
three coefficients).

 For
$Q^2 = (10 \; {\rm GeV})^2$ and $M_0 = 1$ GeV the correction
 in $\alpha_B(Q^2)$ introduced by replacements (\ref{21}) is around 2\% and it
decreases for $Q^2$ growing.

Defining $\Lambda \equiv \Lambda_B = \Lambda_{\overline{MS}}$
(since $\alpha_B$ and $\alpha_s$ practically coincide for $Q^2 >
40$ (GeV)$^2$) one can compare exact $\alpha_B(Q^2)$ obtained by
solving (\ref{8}) with $\alpha_s(Q^2)$ (two loop) of SPT  and
observe a significant difference only at small $Q^2$, $Q^2 \leq
10$ GeV$^2$.

One can see in Table 2 that the difference of  $\alpha_B^{(min)}$
and $\alpha_s^{(2 loop)}$ is 20\% at 1 GeV$^2$ and drops to 10\%
at 5 GeV$^2$, which implies that the use if $\alpha_B^{(min)}$
instead of $\alpha_s$ in the Euclidean region would not produce
much change in the region $Q^2>1$ GeV$^2$ while it improves the
situation for $Q^2\leq 1$ GeV$^2$, where $\alpha_B^{(min)}$
saturates in agreement with experiment and lattice data (see
\cite{1}-\cite{6} and  \cite{14} for the discussion of this
point).

\begin{figure}[!ht]
\begin{picture}(320,210)
\put(0,15){\includegraphics{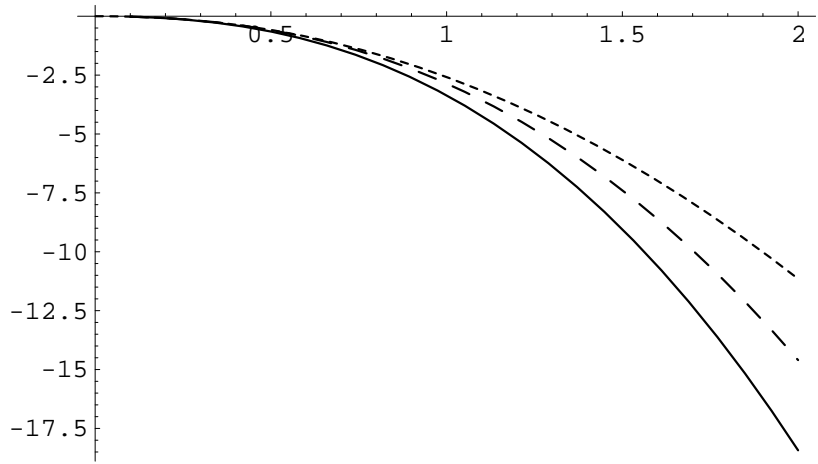}}
\put(0,90){\rotatebox{90}{}}
\put(130,15){}
\put(100,120){}
\end{picture}
\caption{$\beta_{min}(\alpha)$ in the minimal model Eqs.(2), (7) with $P \equiv 0,
\Delta = 1.5$, reproducing scheme-independent coefficients $\beta_0, \beta_1$--
dashed line, and $\beta_{ext}(\alpha)$ in the extended model Eqs.(2), (21) with
$\varepsilon = 3$ and $\Delta = 1.5$ (solid line) and $\Delta = 3$ (dotted line).}
\label{Fig.2}
\end{figure}

6. We have used the large $N_c$ limit to simplify analysis of
analytic properties of background coupling constant $\alpha_B$.
This has enabled us to study $\alpha_B(Q^2)$ also for time-like
$Q^2$ and to suggest a simplified model solution of exact RG
equations, which yields $\alpha_B$ to all orders and prescribes
the nonperturbative expression for $\beta(\alpha)$.

The resulting solution for $\alpha_B$ coincides with the standard
$\alpha_s$ at large $|Q^2|$ and displays familiar AF properties.
At the same time $\alpha_B(Q^2)$ stays finite everywhere in the
Euclidean region, demonstrating phenomenon of IR saturation. For
time-like $Q^2$ the model $\alpha_B(Q^2)$ in the large $N_c$ limit
acquires pole singularities, which have been treated above in the
averaging procedure. It is worth noting that our minimal model
correctly reproduces both scheme-independent coefficients
$\beta_0,\beta_1$. At the same time scheme-dependent coefficients
$\beta_2,\beta_3$ etc. are defined by additional pole terms with
arbitrary constants in them.
 Several improvements are possible with the
suggested model solution. First, one can consider $n_f\neq 0$,
$N_c$ finite, in which case poles would acquire width. This will
not change much the averaged values of $\alpha_B(s)$ calculated
above, but will enable one to make  local predictions of resonance
behaviour. Secondly, our model is too simplified, since
$\psi(\frac{Q^2+M^2_0}{m^2})$ appearing in (\ref{8}), is
responsible only for the equidistant spectrum of hybrids,
(equivalent of gluon loops), but this spectrum should not be
equidistant for lowest states, which yields power corrections,
considered in \cite{10a}.

 The author is grateful to N.O.Agasian and A.V.Radyushkin for helpful discussions; a
 partial financial support of RFFI grants, 00-02-17836,
 00-15-96786 and INTAS grants 00-00110 and 00-00366 are gratefully acknowledged.

 \vspace{1cm}

\begin{center}

\begin{tabular}{|l|l|l|l|l|l|l|l|l|}\hline
$\alpha$&0&0.1&0.2&0.5&1&1.2&1.5&2\\\hline
$\beta_{min}{(\alpha)}$&0&-0.0187&-0.08&-0.587&-2.85&-4.36&-7.38&-14.59\\
\hline $ \beta_{ext}{(\alpha)}$
&0&-0.019&-0.082&-0.65&-3.36&-5.23&-9.04&-18.43\\\hline
$\beta^{(3loop)}_{\overline{MS}}$&0&-0.01895&-0.0833&-0.767&-6.82&-13.56&-33.35&-115.24\\\hline
\end{tabular}
\end{center}
 Table 1:
The QCD $\beta $-function $\beta(\alpha)$ for $n_f=0, N_c=3$
obtained numerically to all orders in $\alpha$ for the minimal
model $\beta_{min}{(\alpha)}$ Eq.(2) with
$\varphi=\frac{\lambda}{\alpha}+\Delta$, and
$\Delta=1.5;~\lambda=1.14$, and for the extended model
$\beta_{ext}$ with $\varphi'$ from (21). The bottom line refers to
$\beta(\alpha)$ in $ \overline{MS}$ as given in Eq. (\ref{4}).

 \vspace{1cm}

\begin{center}

\begin{tabular}{|l|l|l|l|l|l|l|l|l|l|l|}\hline
$Q^2$ (GeV$^2$)&0&0.5&1&1.5&2&3&5&10&40&100\\\hline
$\alpha_B^{(min)}$&0.546&0.394&0.337&0.306&0.286&0.26&0.232&0.201&0.159&0.139\\\hline
$\alpha_B^{(ext)}$&
0.754&0.462&0.379&0.337&0.311&0.278&0.245&0.210&0.163&0.142\\\hline
$\alpha_B^{(2 loop)}$&
0.64&0.359&0.328&0.293&0.27&0.244&0.217&0.188&0.15&0.131\\\hline
$\alpha_s^{(2 loop)}$&
&0.73&0.404&0.33&0.293&0.256&0.229&0.19&0.149&0.131\\\hline\end{tabular}

\end{center}
Table 2: $\alpha_B(Q^2)$ computed numerically to all orders within
minimal model, Eq.(\ref{8}), with $\Lambda=0.37$ GeV,
$\Delta=1.5,~m=M_0=1$ GeV and in extended model, Eqs. (\ref{3}),
(\ref{21}), with $\eta=3, ~ \tilde C=1.253$ reproducing all known
coefficients $\beta_n, n=0,...3$ in $\overline{MS}$ scheme for
$n_f=0$. The last two lines refer to the two-loop approximation of
$\alpha_B$ and $\alpha_s$ (in SPT) with the same parameters.

\end{document}